\documentclass[11pt,a4paper]{article}

\usepackage[english]{babel}
\usepackage[utf8]{inputenc}
\usepackage[T1]{fontenc}
\usepackage{geometry}
\usepackage{setspace}
\usepackage{graphicx}
\usepackage{booktabs}
\usepackage{caption}
\usepackage{subcaption}
\usepackage{amsmath}
\usepackage{url}
\usepackage[none]{hyphenat}
\usepackage{microtype}
\usepackage{ragged2e}
\usepackage[backend=biber, style=apa, citestyle=apa]{biblatex}
\usepackage{xcolor}
\usepackage{tabularx}
\usepackage{authblk}

\justifying

\title{Using Educational Comics in Physics Teaching for Chemistry and Biochemistry Students:\\
Impact on Motivation and Domain-Specific Conceptual Gains}

\author[1,2]{Mauricio Echiburu}
\author[3]{Camilo Henríquez}
\author[4]{Rodrigo Valdés}
\author[5]{Cristóbal Ríos}

\affil[1]{Doctoral Program in Physics, Universidad de La Serena, Chile}
\affil[2]{BSc in Astronomy, Faculty of Engineering and Architecture, Universidad Central, La Serena, Chile}
\affil[3]{Exact Sciences Program, Academic Vice-Rectorate, Universidad Viña del Mar, Viña del Mar, Chile}
\affil[4]{Department of Physics, Universidad Técnica Federico Santa María, Valparaíso, Chile}
\affil[5]{Independent researcher, Chile}

\date{}

\begin{document}
\maketitle

\begin{abstract}
This study analyzes the impact of using educational comics as an active learning strategy in physics workshops for Chemistry and Pharmacy and Biochemistry undergraduate programs during the second semester of 2025. Conceptual understanding was assessed using the Force Concept Inventory (FCI), while motivation and attitudes toward physics were measured through a Likert-type survey administered in pre- and post-test formats.

The results show an average normalized gain of $\langle g \rangle = 0.21$ for the total FCI, corresponding to a low-to-medium range according to physics education research literature. A higher gain is observed in items directly associated with the intervened content ($\langle g \rangle = 0.23$) compared to non-intervened items ($\langle g \rangle = 0.19$), suggesting a relationship between instructional design and domain-specific conceptual development.

At the motivational level, favorable and consistent changes are observed in interest, self-efficacy, and perceived usefulness of the subject, along with a reduction in negative emotional responses toward physics.

These findings suggest that visual narratives can function as an effective pedagogical scaffold, particularly in promoting favorable learning dispositions and supporting conceptual development in specific domains within non-physics university contexts.
\end{abstract}

\section{Introduction}

The teaching of physics in health-related chemical science programs presents persistent difficulties, characterized by low levels of conceptual understanding, limited motivation, and a predominantly instrumental perception of the subject \parencite{Good2019}. In these contexts, physics is often approached as a curricular requirement rather than as a discipline with intrinsic educational value, negatively impacting both conceptual learning and students’ affective engagement \parencite{Leak2018}.

From the perspective of physics education research, these outcomes are consistent with evidence showing that traditional instructional approaches tend to produce limited conceptual gains, particularly in introductory courses for non-specialist students \parencite{Hake1998,Redish2003}. Within this framework, active learning methodologies have been widely proposed as a means to promote greater cognitive engagement and support processes of conceptual change \parencite{Hartikainen2019}.

Among these approaches, the use of visual narratives—and specifically educational comics—has emerged as a pedagogical alternative capable of integrating graphical representation, contextualization, and collaborative discussion \parencite{Ainsworth2006,Lin2015}. From a multimodal learning perspective, such resources facilitate the construction of more robust mental representations and the externalization of conceptual reasoning in less formal environments, potentially contributing both to conceptual learning and to improvements in affective dimensions of the learning process.

In this context, the present study analyzes the impact of the systematic use of educational comics as an active learning strategy in physics workshops within Chemistry and Pharmacy and Biochemistry programs at Universidad Viña del Mar (UVM). The intervention was implemented throughout the second semester of 2025 and involved guided reading, small-group discussion, and the resolution of contextualized activities.

The aim of this study is to evaluate the impact of this intervention on both conceptual understanding and student motivation. To this end, changes in conceptual understanding were assessed using the Force Concept Inventory (FCI) \parencite{Hestenes1992}, while variations in motivation and attitudes toward physics were measured using a Likert-type questionnaire, allowing for an integrated evaluation of cognitive and affective dimensions of learning.

\section{Theoretical Framework}

Physics teaching has been widely studied from the perspective of conceptual learning, showing that mere exposure to content and routine problem-solving do not guarantee a deep understanding of physical phenomena \parencite{Savinainen2013}. In particular, students often enter courses with strongly held prior conceptions that persist even after formal instruction if they are not explicitly addressed. From this perspective, learning is understood as a process of conceptual reorganization rather than simple information accumulation, requiring instructional strategies aimed at confronting, challenging, and reconstructing these initial conceptions \parencite{SavinainenScott2015}.

Within this context, active learning is framed as a pedagogical approach that promotes students’ cognitive engagement through activities involving interpretation, argumentation, and decision-making. The effectiveness of these strategies depends on their alignment with the conceptual domains being addressed and their ability to generate meaningful situations of cognitive conflict \parencite{Hartikainen2019}, that is, situations in which contrasting ideas foster knowledge construction and access to deeper levels of understanding \parencite{Fernandez2022}. Instruments such as the Force Concept Inventory (FCI) \parencite{Hestenes1992,Savinainen2013} are situated within this framework, as they assess conceptual understanding of mechanics from a non-algorithmic perspective, focusing on the consistency of physical reasoning rather than on the application of formulas.

From another perspective, the use of visual narratives in education is grounded in cognitive approaches that emphasize the multimodal nature of learning \parencite{Ainsworth2006}. The combination of text and images supports the construction of more robust mental representations, particularly when addressing abstract or counterintuitive phenomena. In this framework, educational comics constitute a pedagogical resource that integrates narrative sequencing, graphical representation, and contextualization, enabling physical concepts to be situated in contexts closer to students’ experiences \parencite{Lin2015}. This integration facilitates the externalization of reasoning, collective discussion, and the explicit articulation of conceptual errors in a less intimidating environment than traditional formal discourse.

In addition to their cognitive impact, visual narratives may influence affective dimensions of learning \parencite{GarciaCabrero2018}. In programs where physics is not perceived as a core component of professional identity, negative attitudes, low motivation, and high levels of academic anxiety are frequently observed. In particular, emotions such as anxiety, frustration, or rejection toward the subject can negatively affect academic performance \parencite{Villavicencio2011}. In this sense, the introduction of alternative instructional resources, such as comics, may help modify this instrumental perception of the subject, fostering a more positive relationship with both the content and the learning process. From this perspective, motivation is understood not only as an individual trait but as a variable sensitive to the design of learning experiences.

Therefore, the systematic incorporation of comics into physics workshops is justified both from the standpoint of conceptual learning and from the motivational domain. By integrating active learning, visual representation, and collaborative work, this approach aims to create conditions conducive to conceptual change and to a more favorable affective disposition toward the subject. The present study is framed within this perspective, evaluating the impact of this intervention in an integrated manner on conceptual understanding and student motivation, with particular emphasis on its effect on domain-specific conceptual development.

\section{Methodology}

The study followed a quasi-experimental pretest--posttest design without a control group. No parallel pedagogical interventions were implemented during the semester, so the observed results are associated with the described intervention, acknowledging the inherent limitations of this type of design. The course cohort consisted of 60 enrolled students, divided into two workshop sections of 30 students each, taught in parallel schedules.

For the analysis of conceptual understanding, only students who completed both the pre- and post-test were considered, constituting the paired sample analyzed for this indicator. The motivational survey was completed by students present at each administration; the corresponding sample sizes are reported in the Results section.

The pedagogical intervention consisted of the use of educational comics, previously designed as the central activity in physics workshop sessions \parencite{Echiburu2025Comics}. Four one-hour sessions were implemented during weeks 3, 4, 5, 7, and 9 of a total of 16 instructional weeks in the second semester of 2025, organized in collaborative groups of three to four students \parencite{Johnson2014}.

The activities were structured around problem-based situations embedded within the narrative of the comics, which were addressed collaboratively through group discussion and the use of whiteboards. The topics covered through this approach included scientific calculator use, linear kinematics (including graph interpretation), forces, and work--energy. The instructor’s role was that of a learning facilitator, promoting conceptual discussion and collective reflection through guiding questions aimed at activating prior knowledge, generating cognitive conflict, and supporting students’ physical reasoning.

Conceptual understanding was assessed using the 2011 version of the \textit{Force Concept Inventory} (FCI), developed by McCullough and translated into Spanish by Macia-Barber, Hernández, and Menéndez \parencite{PhysPortFCI}. The instrument consists of 30 multiple-choice items and was administered without modifications as a pretest at the beginning of the semester and as a posttest at the end.

Motivation and attitudes toward physics were assessed using a five-point Likert-type survey composed of seven items, specifically designed for this intervention and based on the instrument developed by Cordero et al. \parencite{cordero2017}. The instrument included both positively and negatively worded statements associated with dimensions of interest, self-efficacy, anxiety, and perceived usefulness of physics \parencite{FCIMotivationSurvey2025}.

The study was conducted in accordance with institutional protocols. Participation was voluntary, with informed consent obtained and confidentiality of the data ensured.

\section{Results}

\begin{figure}[ht]
\centering
\includegraphics[width=0.95\linewidth]{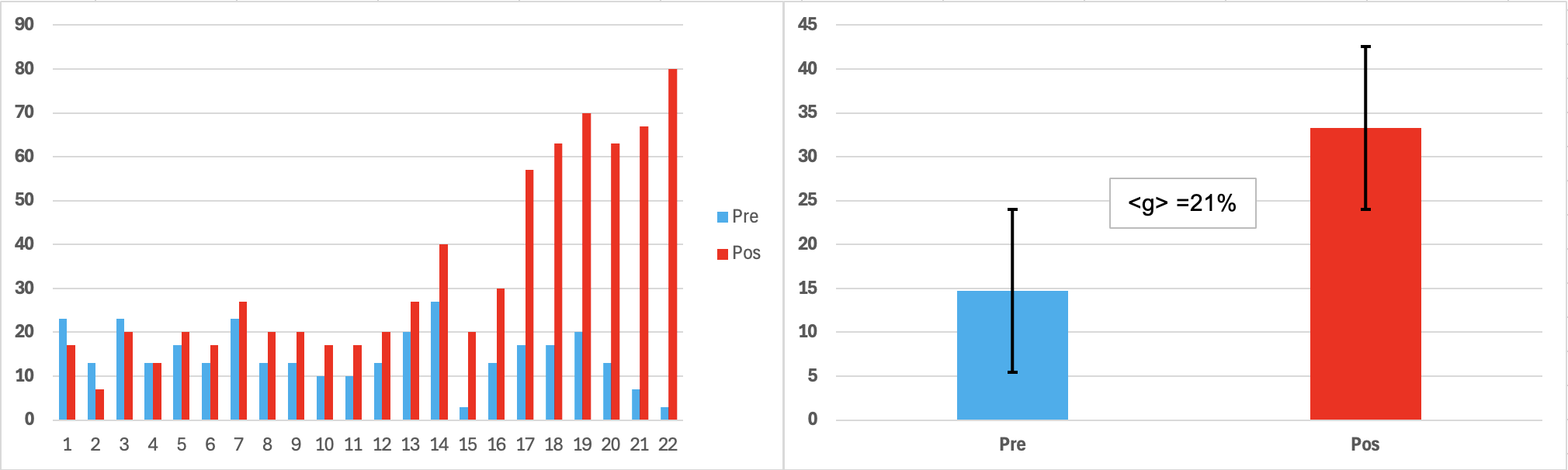}
\caption{(Left) Distribution of Force Concept Inventory scores (30 items) in pretest and posttest for 22 students with paired data. (Right) Average percentage of correct responses in pre and post considering all 30 items and the corresponding normalized gain.}
\label{fig:fci30}
\end{figure}

Figure~\ref{fig:fci30} shows the distribution of Force Concept Inventory (FCI) scores in the pre- and post-test for the 30 items of the instrument for the 22 students with paired data. The left panel presents individual student scores (in percentage) for pre and post, ordered according to the difference between both measurements. The right panel shows the group average percentage for both measurements, together with the normalized gain of 21\% \parencite{Hake1998} calculated over the full set of items.

Considering all 30 FCI items, the average percentage of correct responses increased from approximately 15\% in the pretest to 33\% in the posttest, corresponding to an average normalized gain of $\langle g \rangle \approx 0.21$. This value is consistent with the low-to-medium gain range reported in the literature for introductory courses without extensive curricular transformations.

\begin{table}[ht]
\centering
\caption{Overall results of the Force Concept Inventory (30 items).}
\label{tab:fci30}
\begin{tabular}{lccc}
\hline
Item set & Pretest (\%) & Posttest (\%) & $\langle g \rangle$ \\
\hline
Total FCI (30 items) & 15 & 33 & 0.21 \\
\hline
\end{tabular}
\end{table}

Since the pedagogical intervention focused on specific content domains, a differentiated analysis was conducted by separating FCI items associated with the intervention from those not directly addressed.

For the 20 items directly related to the content addressed through comics, the average percentage of correct responses increased from 12\% in the pretest to 32\% in the posttest, corresponding to a normalized gain of $\langle g \rangle \approx 0.23$. In contrast, for the 10 items not associated with the intervention, the average percentage increased from 20\% to 35\%, corresponding to a normalized gain of $\langle g \rangle \approx 0.19$.

\begin{table}[ht]
\centering
\caption{Force Concept Inventory results by item type.}
\label{tab:fci_tipo}
\begin{tabular}{lccc}
\hline
Item set & Pretest (\%) & Posttest (\%) & $\langle g \rangle$ \\
\hline
Intervention-related items (20) & 12 & 32 & 0.23 \\
Non-intervention items (10) & 20 & 35 & 0.19 \\
\hline
\end{tabular}
\end{table}

\begin{figure}[ht]
\centering
\begin{subfigure}[t]{0.48\linewidth}
\centering
\includegraphics[width=\linewidth]{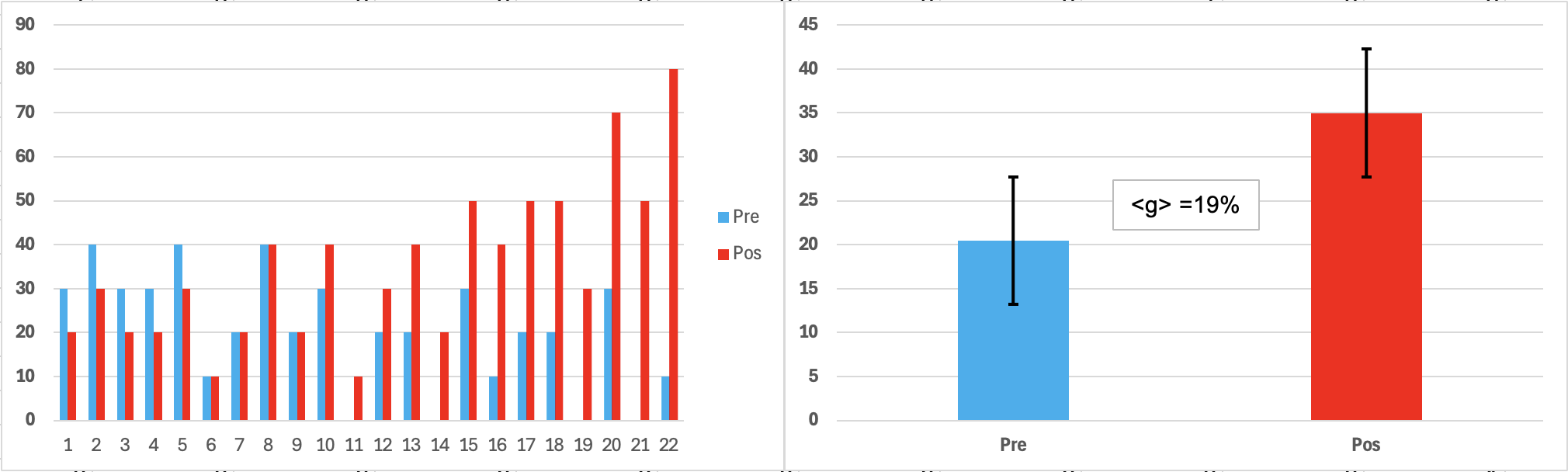}
\caption{Non-intervention items (10 items), $\langle g \rangle = 0.19$.}
\end{subfigure}
\hfill
\begin{subfigure}[t]{0.48\linewidth}
\centering
\includegraphics[width=\linewidth]{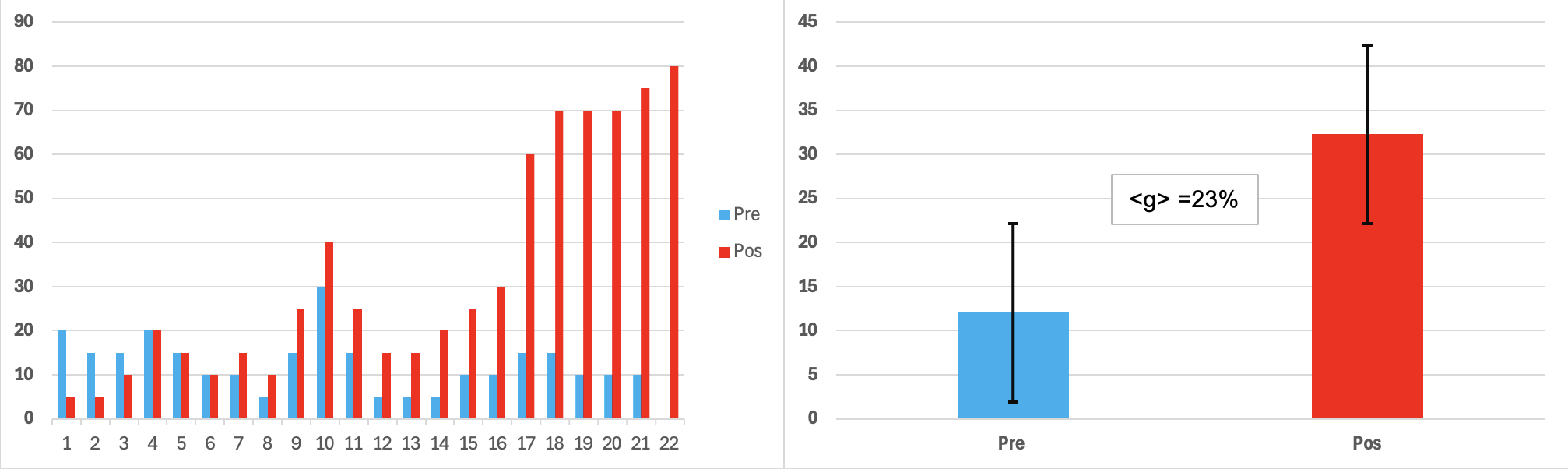}
\caption{Intervention-related items (20 items), $\langle g \rangle = 0.23$.}
\end{subfigure}
\caption{Percentage of correct responses in pre and post by student and group average, separated by item type.}
\label{fig:fci_tipo}
\end{figure}

Figure~\ref{fig:fci_tipo} presents the percentage of correct responses per student in pre and post for both item sets, including the corresponding group averages based on the 22 students with paired data.

For the non-intervention items, the group average increased from approximately 20\% to 35\%, consistent with a gain of $\langle g \rangle = 0.19$ (Figure~\ref{fig:fci_tipo}a). For the intervention-related items, the average increased from 12\% to 32\%, reaching a higher gain of $\langle g \rangle = 0.23$ (Figure~\ref{fig:fci_tipo}b).

The difference observed between both item sets suggests an effect associated with the alignment between the pedagogical intervention and the evaluated conceptual domains. However, the dispersion in individual results indicates a high degree of heterogeneity in learning gains, a common feature in this type of intervention.

Motivation and attitudes toward physics were assessed using a Likert-type survey administered in pre and post formats. The analysis was conducted separately for positively and negatively worded items.

\begin{figure}[ht]
\centering
\includegraphics[width=0.6\linewidth]{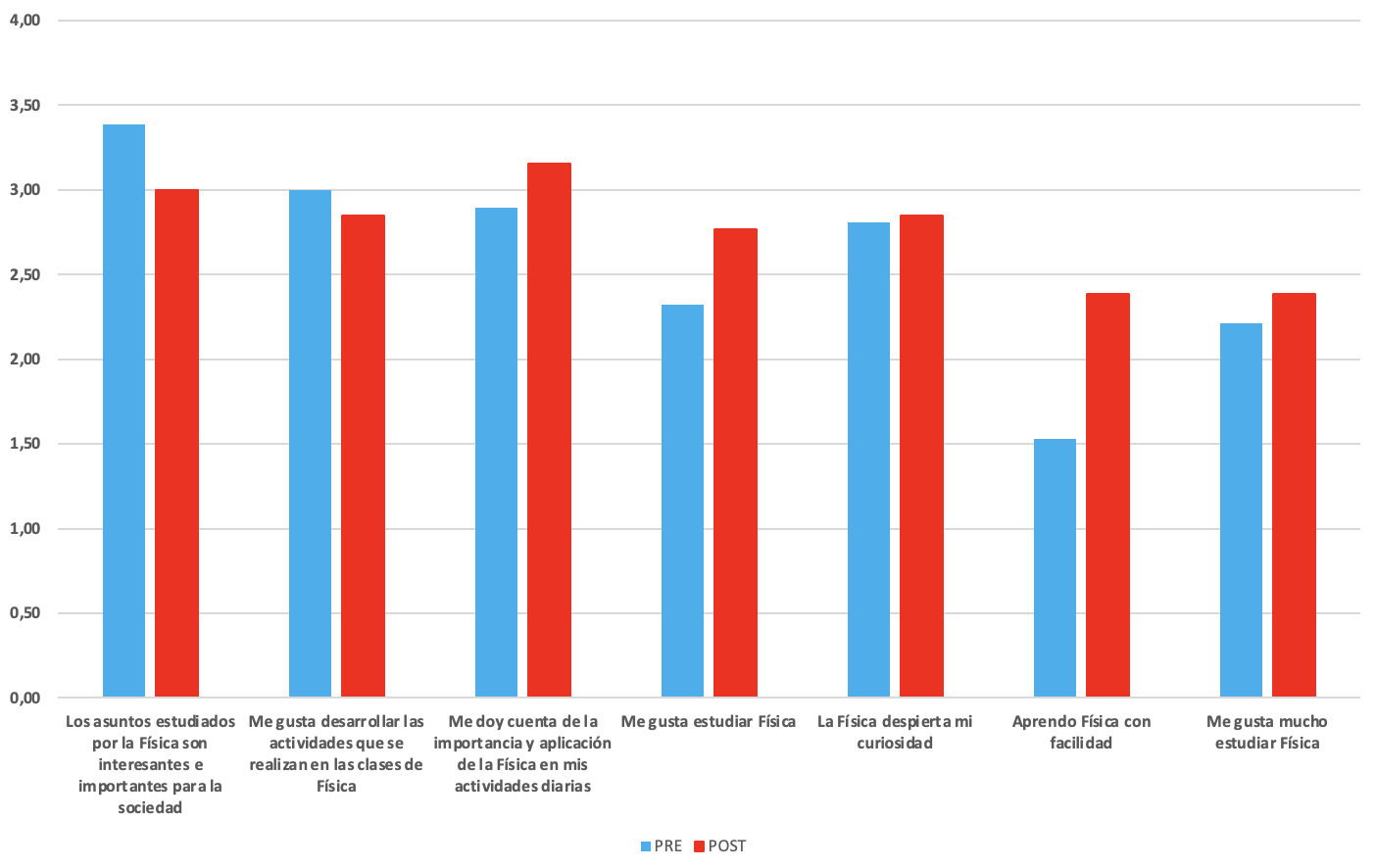}
\caption{Pre and post averages of positively worded items in the motivational survey.}
\label{fig:motiv_pos}
\end{figure}

Figure~\ref{fig:motiv_pos} shows the average scores for the positively worded items before and after the intervention. Overall, increases are observed in most items, suggesting an improvement in students’ disposition toward the subject.

Notable increases are observed in ``I like studying Physics'' (2.32 to 2.77) and ``I learn Physics easily'' (1.53 to 2.38), the latter showing the largest change. Improvements are also observed in items related to perceived usefulness and relevance of the subject.

\begin{table}[ht]
\centering
\caption{Motivational survey results: positively worded items.}
\label{tab:motiv_pos}
\begin{tabularx}{0.95\linewidth}{>{\raggedright\arraybackslash}Xccc}
\hline
Statement & Pre & Post & Change \\
\hline
The topics studied in Physics are interesting and important for society & 3.38 & 3.00 & -0.38 \\
I enjoy the activities carried out in Physics classes & 3.00 & 2.85 & -0.15 \\
I recognize the importance and application of Physics in my daily activities & 2.89 & 3.15 & 0.26 \\
I like studying Physics & 2.32 & 2.77 & 0.45 \\
Physics sparks my curiosity & 2.81 & 2.85 & 0.04 \\
I learn Physics easily & 1.53 & 2.38 & 0.85 \\
I really like studying Physics & 2.21 & 2.38 & 0.17 \\
\hline
\end{tabularx}
\end{table}

\begin{figure}[h]
\centering
\includegraphics[width=0.8\linewidth]{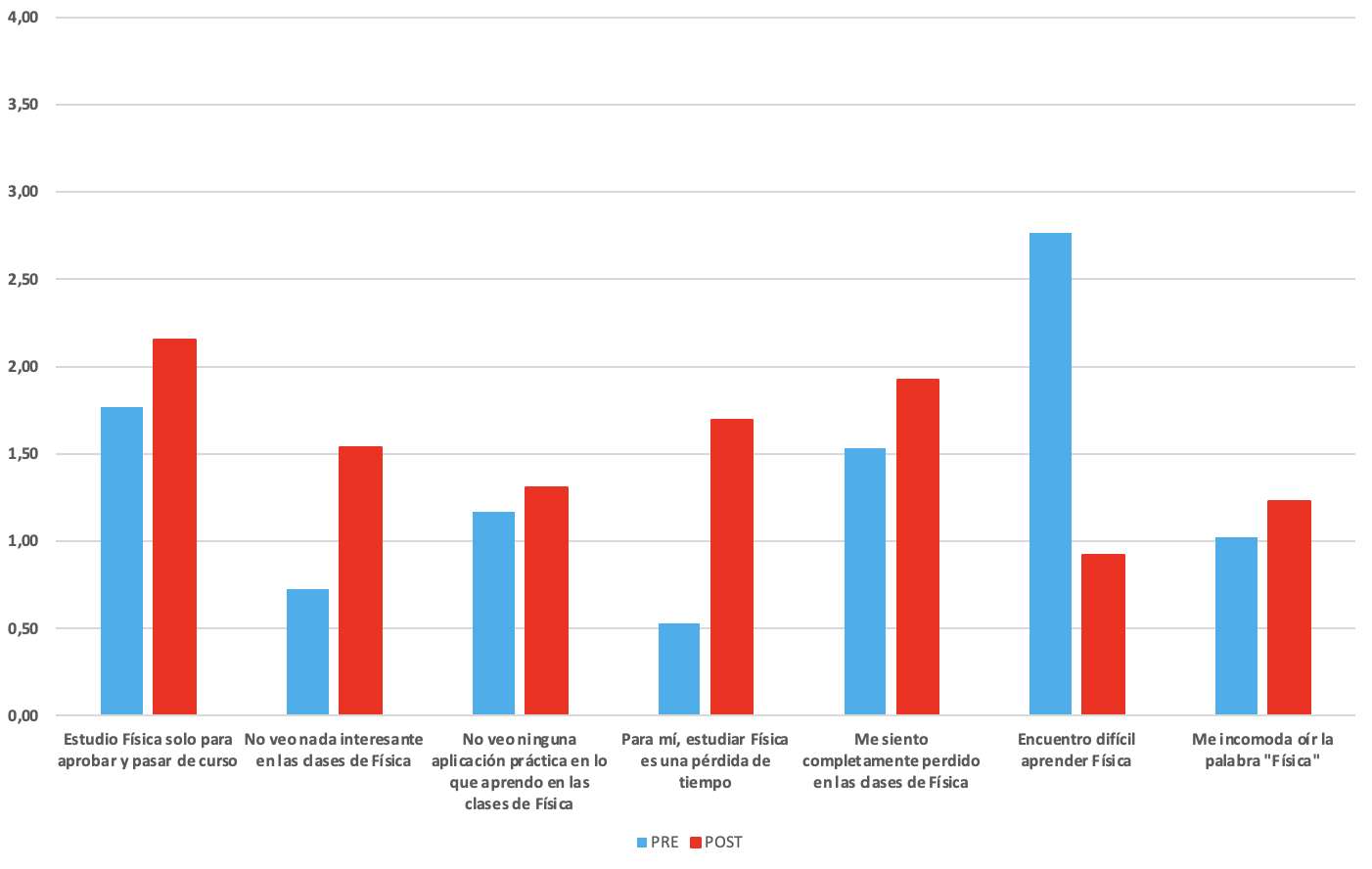}
\caption{Pre and post averages of negatively worded items in the motivational survey.}
\label{fig:motiv_neg}
\end{figure}

The results for negatively worded items are presented in Figure~\ref{fig:motiv_neg}. In general, a decrease in agreement levels is observed in several items, suggesting a reduction in unfavorable perceptions toward the subject.

A substantial decrease is observed in ``I find it difficult to learn Physics'' (2.77 to 0.92), while other items show variations reflecting more heterogeneous responses.

\begin{table}[h]
\centering
\caption{Motivational survey results: negatively worded items.}
\label{tab:motiv_neg}
\begin{tabularx}{0.95\linewidth}{>{\raggedright\arraybackslash}Xccc}
\hline
Statement & Pre & Post & Change \\
\hline
I study Physics only to pass the course & 1.77 & 2.15 & 0.39 \\
I find nothing interesting in Physics classes & 0.72 & 1.54 & 0.82 \\
I see no practical application in what I learn in Physics classes & 1.17 & 1.31 & 0.14 \\
For me, studying Physics is a waste of time & 0.53 & 1.69 & 1.16 \\
I feel completely lost in Physics classes & 1.53 & 1.92 & 0.39 \\
I find it difficult to learn Physics & 2.77 & 0.92 & -1.84 \\
I feel uncomfortable hearing the word Physics & 1.02 & 1.23 & 0.21 \\
\hline
\end{tabularx}
\end{table}

Overall, the results indicate improvements in dimensions related to perceived competence, interest, and perceived usefulness of physics, together with a partial reduction in perceived difficulty. However, the variability observed across some items suggests that the motivational impact is not uniform across all students.

\section{Discussion}

The results obtained indicate a differentiated impact of the comic-based pedagogical intervention on the cognitive and motivational dimensions of physics learning. While the conceptual gains measured through the Force Concept Inventory fall within a low-to-medium range, the changes observed in motivation and attitudes toward the subject are more consistent.

In terms of conceptual understanding, the average normalized gain $\langle g \rangle \approx 0.21$ for the full FCI is consistent with previous studies in non-physics university contexts, where introductory courses typically show limited improvements in the absence of substantial curricular transformations \parencite{SavinainenScott2015}. In this sense, the results align with expected outcomes for targeted and short-duration interventions.

However, the item-level analysis reveals a particularly relevant result: items directly associated with the content addressed through comics show a higher gain ($\langle g \rangle \approx 0.23$) than those not directly targeted ($\langle g \rangle \approx 0.19$). This finding suggests that the effectiveness of the intervention is not uniform, but rather depends on the alignment between instructional design and the evaluated conceptual domains, in agreement with findings reported in the active learning literature \parencite{ValdesGuajardo2025}.

The heterogeneity observed in individual gains, for both intervention-related and non-intervention items, reinforces the idea that such strategies do not impact all students equally. This behavior has been widely documented in the literature, where factors such as prior conceptions, level of engagement in collaborative work, and individual disposition toward the subject significantly influence learning outcomes \parencite{RodriguezBorges2020}.

In contrast, the results from the attitudinal instrument show more pronounced changes. Positive items exhibit increases in dimensions related to self-efficacy and interest in the subject, while some negative items show a decrease in perceived difficulty. These findings suggest that the intervention contributes to modifying students’ affective relationship with physics, reducing emotional barriers that often interfere with conceptual learning \parencite{mosquera2024}.

However, certain items associated with perceived relevance of the subject, such as ``I find nothing interesting in Physics classes'' and ``For me, studying Physics is a waste of time'', show an increase in agreement levels. This seemingly contradictory result can be interpreted in light of the curricular context in which the course is situated. In particular, the isolated role of physics within these programs may limit its perceived professional relevance, regardless of improvements in the classroom learning experience.

Overall, the observed asymmetry between moderate conceptual gains and more consistent motivational changes suggests that the primary contribution of comics lies initially in enhancing students’ disposition toward learning. From this perspective, visual narratives can be understood as a pedagogical scaffold that creates more favorable conditions for conceptual change, rather than as a direct mechanism for conceptual improvement in itself.

\section{Conclusions}

This study shows that the systematic incorporation of educational comics as the central activity in physics workshops for Chemistry and Pharmacy and Biochemistry programs is associated with favorable changes in student motivation and with moderate conceptual gains, particularly in the domains directly addressed by the intervention.

The results suggest that visual narratives constitute a relevant pedagogical resource in non-physics university contexts, where the subject is often perceived in an instrumental way. In this scenario, their use promotes student participation, reduces affective barriers, and strengthens perceived competence, all of which are key elements for sustaining learning processes.

From a conceptual perspective, although the observed gains fall within a low-to-medium range, the higher gains in items aligned with the intervention highlight the importance of coherence between instructional design and the targeted conceptual domains. This finding suggests that the use of comics may contribute specifically to conceptual development when explicitly integrated with learning objectives.

Overall, the results indicate that the primary contribution of this strategy lies in creating more favorable conditions for learning, rather than in directly increasing overall conceptual performance.

Future research could explore longer interventions, incorporate designs with control groups, and include follow-up assessments in order to better understand the long-term impact of this type of strategy on conceptual understanding in physics.

\section*{Acknowledgments}

The authors acknowledge the financial support provided by the Fondo de Creación e Investigación Formativa (FCIF) 2025 of Universidad Viña del Mar, which made this research possible.

\printbibliography

\end{document}